\documentclass[%
 reprint,
 amsmath,amssymb,
 aps,
]{revtex4-1}

\usepackage{graphicx}
\usepackage{dcolumn}
\usepackage{bm}

\begin{document}

\preprint{APS/123-QED}

\title{Properties of protons in nuclear medium with AdS/QCD model with a quadratic modified dilaton}

\author{Alfredo Vega}
\email{alfredo.vega@uv.cl}

\author{M. A. Martin Contreras}%
\email{miguelangel.martin@uv.cl}
\affiliation{%
 Instituto de F\'isica y Astronom\'ia, \\
 Universidad de Valpara\'iso,\\
 A. Gran Breta\~na 1111, Valpara\'iso, Chile
}%

\date{\today}

\begin{abstract}
A modified version of the usual soft wall quadratic dilaton, a dilaton which catches the variation of the proton mass inside nucleons in the nuclear medium is used. With this dilaton, called \emph{nuclear dilaton}, we have calculated the electromagnetic form factors and magnetic moments associated with the in-medium proton following the AdS/QCD prescription.  
\begin{description}

\item[PACS numbers]
13.40.Gp, 11.25.Tq

\end{description}
\end{abstract}

\maketitle


\section{\label{sec:level1}Introduction}

It is well-known that hadronic properties change inside the nuclear medium \cite{Saito:2005rv}. Although studying these properties should consider QCD, its non-perturbative nature tends to make the calculations difficult. This has motivated the development of several tools to address these inquiries, such as lattice QCD (e.g., \cite{Aoki:2013ldr}), Schwinger-Dyson formulation (e.g., \cite{Roberts:1994dr}), and the formulation of the phenomenological models that capture some of the hadronic constituent properties and make it possible to perform calculations.  Among such models, we can distinguish those based on the AdS/CFT correspondence \cite{Maldacena:1997re, Maldacena:1997re, Witten:1998qj, Witten:1998zw, Gubser:1998bc}.

The holographic dual of QCD is currently unknown, but we can use the AdS/CFT idea to build bottom-up phenomenological models that simulate some of the most important attributes of non-perturbative QCD as well as hadronic physics. These approaches are called \emph{AdS/QCD models}. A close look at the literature reveals that AdS/QCD models have been successfully applied to a wide range of hadronic phenomenology. For example, hadrons in vacuum hadrons at zero  temperature (e.g., see \cite{Karch:2006pv, Kwee:2007dd, Brodsky:2007hb, Vega:2010ns, Branz:2010ub, Braga:2011wa, Vega:2012iz, Watanabe:2013spa, Gutsche:2015xva, Capossoli:2015ywa, Watanabe:2015mia, Capossoli:2016ydo, Gutsche:2017oro, Gutsche:2017lyu}), at finite dense (e.g., see \cite{Rozali:2007rx, Kim:2007xi, Kim:2009ey, Park:2011ab, Sachan:2011iy, Lee:2014gna, Lee:2014xda}) and at finite temperature hadrons (e.g., see \cite{Fujita:2009wc, Miranda:2009qp, Colangelo:2013ila, Bellantuono:2014lra, Braga:2016wkm, Vega:2017dbt, Vega:2018dgk}). In these models, in-medium properties are enclosed by the geometric background where the excitations, dual to the hadrons of interest, live. 

From the gravitational point of view, the metric and the dilaton field define the background. Both of them are related dynamically as solutions of the equations of motion obtained from the Hilbert-Einstein action variations \cite{Gursoy:2007cb, Gursoy:2007er, dePaula:2008fp, Yang:2015aia, Li:2011hp, He:2013qq, Zollner:2018uep}, but in the holographic of a phenomenological AdS/QCD-like context, it is common to use static dilatons (which are not related to the metric) (e.g., see \cite{Karch:2006pv, Kwee:2007dd, Brodsky:2007hb, Vega:2010ns, Branz:2010ub, Braga:2011wa, Vega:2012iz, Watanabe:2013spa, Gutsche:2015xva, Capossoli:2015ywa, Watanabe:2015mia, Capossoli:2016ydo, Gutsche:2017oro, Gutsche:2017lyu}) to softly break the conformal symmetry, but such a field is not considered part of the background able to catch medium properties that affect the hadrons studied in these models. 

The approximation that considers the nuclear medium effects on the hadronic properties usually includes additional fields in the bulk, the influence of the nucleus on a hadron in such a medium to be adapted holographically.\cite{Kim:2007em}. This last idea leads to a specific metric called \emph{thermal charged AdS} (tcAdS)  \cite{Lee:2009bya, Park:2009nb, Jo:2009xr}. Using this proposal it was possible to address some hadronic properties \cite{Lee:2013oya, Hong:2006ta, Lee:2014xda, Lee:2014gna, Lee:2015rxa, Colangelo:2012jy, Mamedov:2016ype}.

The approach developed here is different, and it is motivated  by three phenomenological aspects that can be implemented simply and consistently in an AdS/QCD model with a quadratic dilaton if one opens to the possibility of, as being part of the background, the dilaton can capture properties of the medium and complement the information enclosed in the metric. First, by looking at the QCD phase diagram, at low temperatures hadrons in vacuum and hadrons inside nucleus are in the same region \cite{Baym:2017whm}. This implies that the holographic formulation of in-medium hadrons could be addressed by using asymptotically AdS spaces, as the ones used in AdS/QCD soft wall-like models, and properties of the medium could be enclosed in the other background ingredient at hand: the dilaton field.  In a previous work, we considered this idea in the context of finite temperature (that we called \emph{thermal dilaton}), that makes it possible to study the melting temperature of hadrons inside a thermal bath \cite{Vega:2018eiq}. This approximation could be useful in upgrading the AdS/QCD models at finite temperature.

Second, hadron masses, which in medium vary with the medium density according to specific scaling rules \cite{Saito:2005rv}, can be implemented easily in the quadratic dilaton context and by considering a small change in the scalar field used to break the conformal invariance.  

And third, the currents that define the nucleon form factors, in vacuum or in medium, with the same structure \cite{Ramalho:2012pu}. This suggests that on the AdS side, the interaction terms that give rise to the form factors also have the same structure. Therefore, the expressions for the form factors in terms of the bulk fields should be similar. 

In this work, following the previous discussion,  we propose a simple extension of the quadratic dilaton, altogether with the AdS/QCD model used to study the proton electromagnetic properties. As in many of the soft wall models, we consider a static dilaton whose functional form is chosen to reproduce the mass spectrum observed. In this case, it follows the scaling rule given in \cite{Saito:2005rv}, which in the case of the holographic models with a quadratic dilaton, is easy to implement. To do so, it is enough to change the parameter used to define the kind of dilaton, giving as a result a dilaton that depends on the density, that we call a \emph{nuclear dilaton}. This new dilaton captures some of the properties of the nuclear medium we are interested in. With this dilaton, we used the results reported in \cite{Gutsche:2012bp, Gutsche:2012wb} to study electromagnetic properties of nucleons, obtaining good results.

This work is structured as follows: in section II we show the phenomenological framework that led us to the modifications of the holographic models, namely the QCD phase diagram, the scaling rules and the proton form factors. In section III we discuss how to modify the soft wall model with a quadratic dilaton according to the phenomenology discussed in section II. Finally, in section IV we give conclusions and final remarks about this work.




\section{Phenomenological Framework}

In this section, we will show the phenomenological background that motivates us to suggest how the dilaton can be modified to enclose the medium properties where the hadron of interest lives.  We will focus on protons. 


\subsection{Nucleons and QCD Phase diagram}

By observing the QCD phase  diagram\cite{Baym:2017whm}, it is possible to note that, at low temperatures and densities close to the nuclear medium,  nucleons live in this region, and do not experience any phase transition.

These ideas led us to explore the possibility of studying hadrons in the nuclear medium with the AdS/QCD  models, where the metric used to study them is the same as the one employed  in vacuum, with the extra ingredient that the nuclear medium properties can be captured in another background field, the dilaton.  

\subsection{Hadron masses in nuclei}

The hadron mass ($M^*$) inside the atomic nucleus varies according to the scaling rule, that in the case of the proton, takes the following form:

\begin{equation}
\label{MasaNucleonMedio}
M^{*} =\left(1 - 0.21 \,\frac{\rho_{B}}{\rho_{0}}\right) M,
\end{equation}
where $M$ is the mass in the vacuum, $\rho_B$ is the medium density and $\rho_0$ is the nucleus density.

This scaling in hadron masses can be implemented in a straightforward way in AdS/QCD models with quadratic dilaton. 

\subsection{Electromagnetic form factors of nucleons inside the nucleus}

Nucleon electromagnetic form factors $F_1^N$ and $F_2^N$ 
($N=p, n$ labels proton and neutron) are conventionally 
defined by the matrix element of the electromagnetic current $J^\mu_\text{EM}$ as

\begin{equation}
\langle p' | J^{\mu}_\text{EM}(0)  | p \rangle = \bar{u}(p')[ \gamma^{\mu} F_{1}^N(Q^{2}) 
+ \frac{i \sigma^{\mu \nu}}{2 m_N}   q_\nu 
F_{2}^N(Q^{2})] u(p),
\end{equation}
where $q = p' - p$ is the momentum transferred, 
$m_N$ is the nucleon mass,
$F_1^N$ and $F_2^N$ are the Dirac and Pauli form factors, which 
are normalized to the electric charge $e_N$ and the anomalous 
magnetic moment $k_N$ of the corresponding nucleon: 
$F_1^N\left(0\right)=e_N$ and $F_2^N\left(0\right)=k_N$.

With the Dirac and Pauli form factors it is possible to build the so-called Sach electric and magnetic form factors as

\begin{equation}
G_{E}\left(Q^{2}\right) = F_{1}\left(Q^{2}\right) - \frac{Q^{2}}{4\, M^{2}}\, F_{2}\left(Q^{2}\right)
\end{equation}
and
\begin{equation}
G_{M}\left(Q^{2}\right) = F_{1}\left(Q^{2}\right) + F_{2}\left(Q^{2}\right).
\end{equation}

At $Q^{2}=0$ the last Sach form factor defines the magnetic moment in nuclear magneton units.

In order to consider medium effects, we use an asterisk to establish a difference with quantities in the vacuum. Assuming that a baryon is quasi-free in the nuclear medium, the electromagnetic current for protons can be expressed as \cite{Ramalho:2012pu} 

\begin{equation}
\langle p' | J^{\mu}_\text{EM}(0) | p \rangle = 
\bar{u}(p') [ \gamma^{\mu} F_{1}^{N*}(Q^{2}) 
+ \frac{i \sigma^{\mu \nu}}{2\,m_N^{*}}  \, q_\nu F_{2}^{N*}(Q^{2})] u(p),
\end{equation}
where $F_1^{N*}$ and $F_2^{N*}$ are the Dirac and Pauli form factors in nuclear medium, which are normalized to electric charge $e_N$ and anomalous magnetic moment $k_N$ of the corresponding nucleon: 
$F_1^{N*}\left(0\right)=e_N$ and $F_2^{N*}\left(0\right)=k_N^{*}$.

As in the vacuum, we define the electric and magnetic form factors as 
\begin{equation}
G_{E}^{*}\left(Q^{2}\right) = F_{1}^{*}\left(Q^{2}\right) - \frac{Q^{2}}{4 \left(M^{*}\right)^{2}}\, F_{2}^{*}\left(Q^{2}\right)
\end{equation}
and
\begin{equation}
G_{M}^{*}\left(Q^{2}\right) = \left[ F_{1}^{*}\left(Q^{2}\right) + F_{2}^{*}\left(Q^{2}\right) \right]\frac{M}{M^{*}},
\end{equation}
where in the last expression an additional factor was included in order to translate the magnetic moments into nuclear magneton units to more easily compare with the  vacuum results.

The structure of the currents that define the electromagnetic form factors is the same in both cases. On the other hand, in soft wall-like models, the interaction terms that give rise to such electromagnetic form factors should have the same structure also. Therefore, these observations make it possible to adapt the existent finite temperature and zero density models to perform calculations for the hadronic properties in nuclear medium. 




\section{Soft wall-like model for nuclear medium}

In AdS/QCD soft wall-like models with an AdS metric and a quadratic dilaton, the nucleon spectrum is calculated as follows  \cite{Gutsche:2012bp}:

\begin{equation}
\label{MasaNucleonVacio}
M_{n}^{2} = 4 \,\kappa^{2} \left(n + 2\right).
\end{equation}

On the other hand, as we mentioned above, the nucleon mass $M^*$ inside a nucleon varies according to \cite{Saito:2005rv}

\begin{equation}
M^{*} = \left(1 - 0.21\, \frac{\rho_{B}}{\rho_{0}}\right) M.
\end{equation}

By comparing these two equations, we infer that the in-medium  nucleon mass can be written as  


\begin{equation}
\label{MasaNucleonMedioHolo}
M_{n}^{*2} = 4\, \kappa^{*2} \left(n + 2\right),
\end{equation}

where

\begin{equation}
\label{kappaN}
\kappa^{*} = \sqrt{\left(1 - 0.21\,l \frac{\rho_{B}}{\rho_{0}}\right)} \,\kappa.
\end{equation}

The latter suggests to us that the effect of the rest of the nucleus on the hadron of interest can be addressed in a soft wall-like model using  a different definition of the dilaton field worked to study hadrons in the vacuum, that we call \emph{nuclear dilaton}, since it captures some properties of the nuclear medium where the hadron is living.

Considering that the hadrons in which we are interested and those in the vacuum are close in the QCD phase diagram, and given that there is no phase transitions, in other words, we are still in the confined phase, we do not change the AdS metric to an AdS-Schwartzchild black hole \cite{Herzog:2006ra}. This allows us to use the soft wall results at zero temperature and zero density with a slight modification to obtain the hadronic properties of hadrons in atomic nuclei. In our case, we will use the approach discussed in  \cite{Gutsche:2012bp} to calculate properties of protons in medium.  Therefore, using (\ref{kappaN}) the mass turns out to be (\ref{MasaNucleonMedio}). Thus, we can calculate the electromagnetic form factors using the expressions 

\begin{equation}
F_{1}^{*}\left(Q^{2}\right)=C_{1}^{*}\left(Q^{2}\right) + g_{v}\, C_{2}^{*}\left(Q^{2}\right) + \eta_{V}^{p}\, C_{3}^{*}\left(Q^{2}\right)
\end{equation}
and 
\begin{equation}
F_{2}^{*}\left(Q^{2}\right)=\eta_{V}^{p}\, C_{4}^{*}\left(Q^{2}\right).
\end{equation}

In this work we must take into account the contribution coming from the valence part, so we used the expressions given in \cite{Gutsche:2012bp}, considering $\tau = 3$. This leads us to functions $C_{i}^{*}\left(Q^{2}\right)$, which are given by

\begin{equation}
C_{1}^{*}(Q^{2})=\frac{1}{2} \int{dz\, V^*(Q,z)\left\{[f_{L}^{*}(z)]^{2} + [f_{R}^{*}(z)]^{2}\right\}}
\end{equation}

\begin{equation}
C_{2}^{*}(Q^{2})=\frac{1}{2} \int{ dz\, V^{*}(Q,z)\left\{[f_{L}^{*}(z)]^{2} - [f_{R}^{*}(z)]^{2}\right\}}
\end{equation}

\begin{equation}
C_{3}^{*}(Q^{2})=\frac{1}{2} \int dz\,z \partial_{z}V^{*}(Q,z)\left\{[f_{L}^{*}(z)]^{2} - [f_{R}^{*}(z)]^{2}\right\}
\end{equation}

\begin{equation}
C_{4}^{*}(Q^{2})= 2 M^{*} \int{ dz \,z\,V^{*}(Q,z)\left\{[f_{L}^{*}(z)]^{2}\,[f_{R}^{*}(z)]^{2}\right\}},
\end{equation}
where we used the $\kappa^*$ parameter for the nuclear medium to calculate $V^{*}\left(Q,z\right)$ and $f_{L/R}^{*}\left(z\right)$. 

By considering the part with $\tau=3$ only, some of the parameters are changed with respect to those used in \cite{Gutsche:2012bp}, and throghoutt this paper we take $g_{v}=0.3$, $\eta_{V}^{p}=0.224$ and $\eta_{V}^{p}=-0.239$.

Following the above discussion, we calculated the electromagnetic form factors and the magnetic moment of the proton inside a nucleus. These results are displayed in Figures I and II. 

\begin{center}
\begin{figure*}
  \begin{tabular}{c c}
    \includegraphics[width=3.4 in]{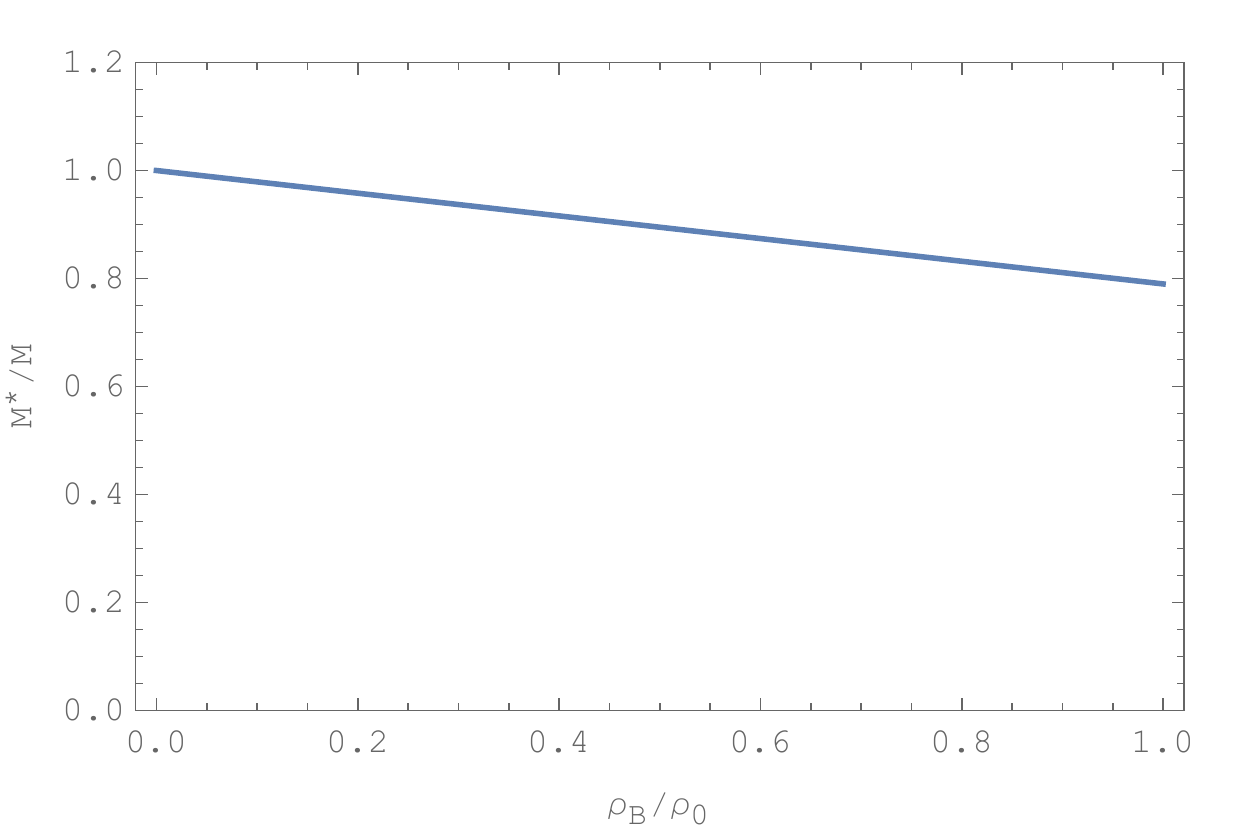}
    \includegraphics[width=3.4 in]{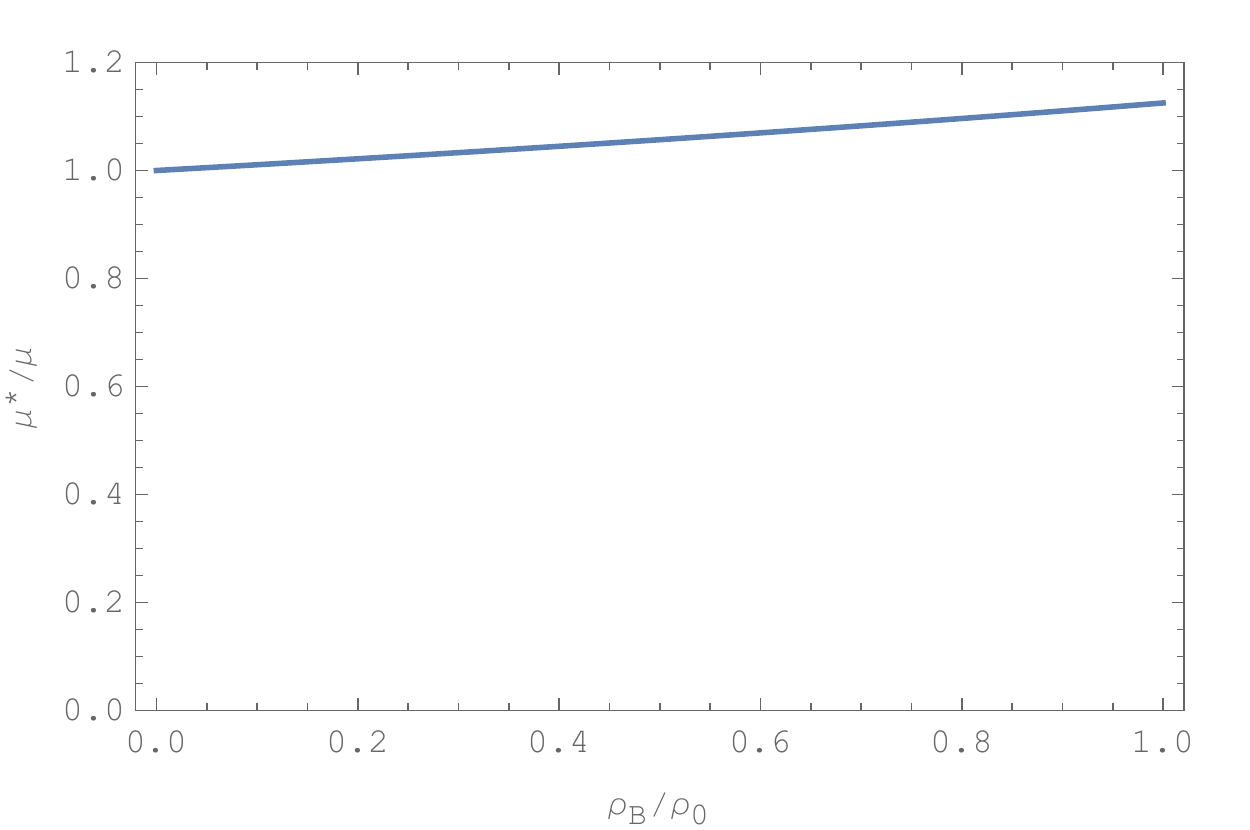}
  \end{tabular}
\caption{The left panel shows the quotient between the mass in medium and the mass in vacuum. The right panel shows the magnetic moment as function of the density in medium divided by the nuclear density.}
\end{figure*}
\end{center}

\begin{center}
\begin{figure*}
  \begin{tabular}{c c}
    \includegraphics[width=3.4 in]{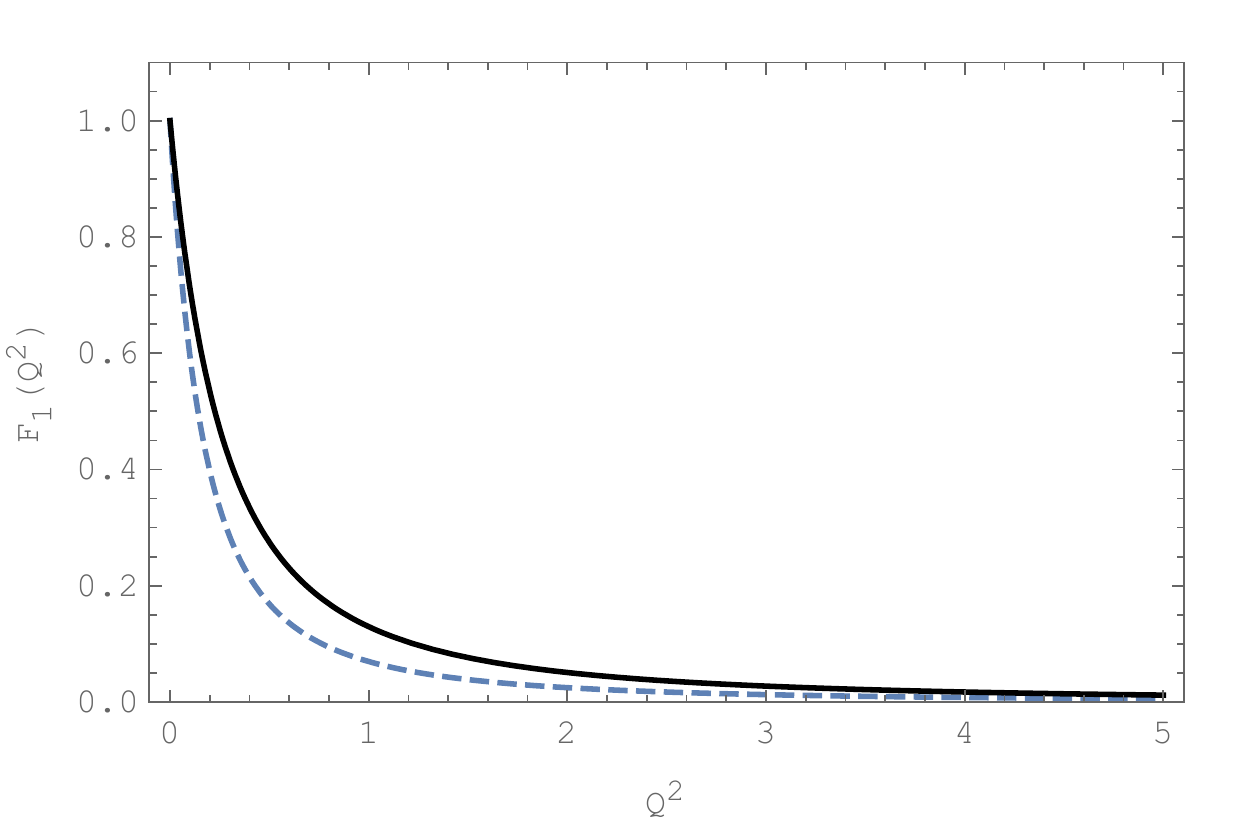}
    \includegraphics[width=3.4 in]{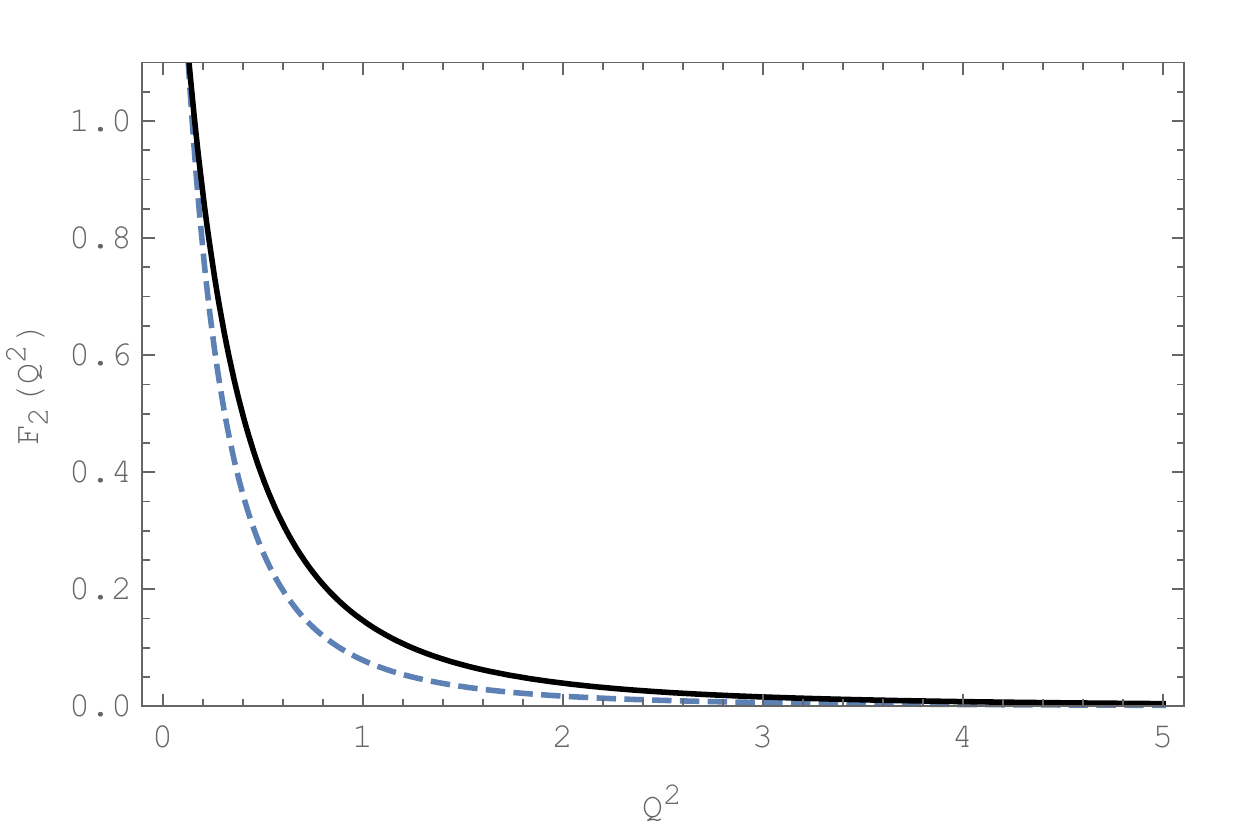}
  \end{tabular}
\caption{The graphics show the Dirac and Pauli form factors for protons when $\rho_{B}/\rho_{0} = 0$ (continuos line) and $\rho_{B}/\rho_{0} = 1$ (dashed line).}
\end{figure*}
\end{center}




\section{Conclusions}

In AdS/QCD models, the properties of hadrons in dense media or finite temperature, in general, consider dealing with AdS black hole (AdS-BH) geometries. In the case of temperature only, we have two possible solutions: thermal AdS for the confined phase and AdS-Schwartzchild black hole in the deconfined phase \cite{Herzog:2006ra}, and there is also the \emph{themal charged AdS} (tcAdS)  \cite{Lee:2009bya, Park:2009nb, Jo:2009xr}, which is sometimes used to calculate properties of hadrons in nuclear media. In all of these cases, medium properties are coded exclusively in metric.

Although dilatons used in most AdS/QCD models are static, it should be remembered that there are background fields (which in dynamical models are related to metric) and as part of the background we suggest they could be a complement to metric, capturing part of the medium properties, and in this way they can help describe hadron properties in medium and/or finite temperature.

To the best of our knowledge, the only attemp to incorporate part of medium properties in a dilaton was done in \cite{Vega:2018eiq}, where the authors consider a thermal dilaton, i.e., a dilaton that depends on temperature also, and they show with this it is possible to modify melting hadron temperatures.

In this paper, we explore the same idea to study proton properties inside the atomic nucleus, and we consider a modified dilaton taking into account three phenomenological aspects that suggest a simple modification in AdS/QCD models with an AdS metric and a quadratic dilaton. As illustrated in Figures I and II, this yields good quantitative results. The hadron mass is increased with nuclear density, magnetic moments are reduced and the electromagnetic form factors follows a similar behavior obtained with other approaches. 

In future studies, we plan to delve more deeply into other proton properties in the nucleus with the extension suggested in this paper, and also to study properties of other hadrons.

\vspace{0.2cm}
\noindent
\textbf{Acknowledgements: } The authors acknowledge the financial support of FONDECYT (Chile) under Grants No. 1180753 (A. V) and No. 3180592 (M. A. M).

\end{document}